# Wakefields and Instabilities in Linear Accelerators

*M. Ferrario, M. Migliorati, and L. Palumbo*
INFN-LNF and Università di Roma 'La Sapienza'

**Abstract**
When a charged particle travels across the vacuum chamber of an accelerator, it induces electromagnetic fields, which are left mainly behind the generating particle. These electromagnetic fields act back on the beam and influence its motion. Such an interaction of the beam with its surroundings results in beam energy losses, alters the shape of the bunches, and shifts the betatron and synchrotron frequencies. At high beam current the fields can even lead to instabilities, thus limiting the performance of the accelerator in terms of beam quality and current intensity. We discuss in this lecture the general features of the electromagnetic fields, introducing the concepts of wakefields and giving a few simple examples in cylindrical geometry. We then show the effect of the wakefields on the dynamics of a beam in a linac, dealing in particular with the beam breakup instability and how to cure it.

## 1   Introduction

Self-induced electromagnetic (e.m.) forces in an accelerator are generated by a charged particle beam which interacts with all the components of the vacuum chamber. These components may have a complex geometry: kickers, bellows, RF cavities, diagnostics components, special devices, etc. The study of the fields generally requires the solution of Maxwell's equations in a given structure, taking the beam current as the source of the fields. This could be a quite complicated task, and therefore several dedicated computer codes, used to study and design accelerator devices, which solve the e.m. problem in the frequency or time domain, have been developed. These include, for example, CST Studio Suite [1], GDFIDL [2], ACE3P [3], ABCI [4], and others.

In this lecture, we discuss some general features of the self-induced e.m. forces and introduce the concepts of wakefields and coupling impedances [5–12], and we present some simple examples in cylindrical geometry. Although the space charge forces have been studied separately [13], they can be seen as a particular case of wakefields.

In the second part of the lecture we study the effects of the wakefields on the dynamics of a beam in a linac, such as energy loss and energy spread. Finally, we deal with the beam breakup (BBU) instability [14], and the way to cure it [15].

## 2   Wake fields and potentials

### 2.1   Longitudinal and transverse wakefields

The self-induced e.m. fields acting on a particle inside a beam depend on the whole charge distribution. However, if we know the fields in a given structure that are created by a single charge (i.e. we obtain the Green function of the structure), by using the superposition principle we can easily reconstruct the fields produced by any charge distribution.

The e.m. fields created by a point charge act back on the charge itself, and on any other charge of the beam. Referring to the coordinate system of Fig. 1, let us call $q_0(s_0, \boldsymbol{r}_0)$ the source charge, travelling with constant longitudinal velocity $v = c$ (ultrarelativistic limit) along a trajectory parallel to the axis of a given accelerator structure. Let us consider a test charge $q$, in a position $(s = s_0 - z, \boldsymbol{r})$, which is moving with the same constant velocity on a parallel trajectory inside the structure.

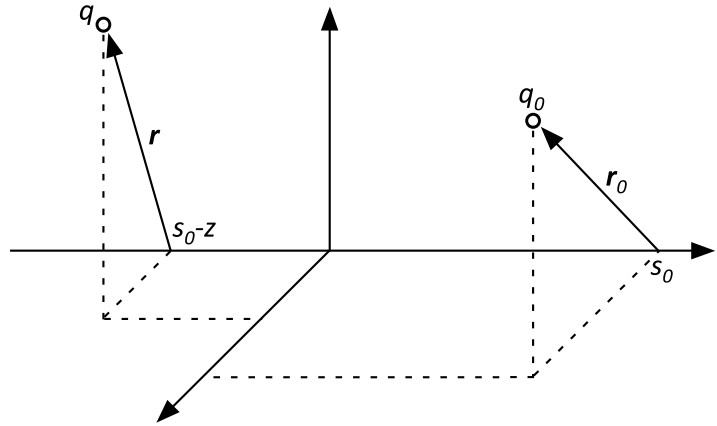

**Fig. 1:** Reference coordinate system

Let $\boldsymbol{E}$ and $\boldsymbol{B}$ be the electric and magnetic fields generated by $q_0$ inside the structure. Since the velocity of both charges is along $z$, the Lorentz force acting on $q$ has the following components:

$$\boldsymbol{F} = q\left[ E_z \hat{z} + \left(E_x - vB_y\right)\hat{x} + \left(E_y + vB_x\right)\hat{y} \right] \equiv \boldsymbol{F}_{/\!/} + \boldsymbol{F}_\perp . \tag{1}$$

From Eq. (1) we see that there can be two effects on the test charge: a longitudinal force, which changes its energy, and a transverse force, which deflects its trajectory. If we consider a device of length $L$, the energy change (in joule) of $q$ due to this force is given by

$$U(\boldsymbol{r},\boldsymbol{r}_0,z) = \int_0^L F_{/\!/} \, \mathrm{d}s , \tag{2}$$

and the transverse deflecting kick (expressed in Newton metres), is given by

$$\boldsymbol{M}(\boldsymbol{r},\boldsymbol{r}_0,z) = \int_0^L \boldsymbol{F}_\perp \, \mathrm{d}s . \tag{3}$$

Note that the integration is performed over a given path of the trajectory. The quantities given by Eqs. (2) and (3), normalized to the two charges $q_0$ and $q$, are called the longitudinal and transverse *wakefields*, respectively. In many cases, we deal with structures having particular symmetric shapes, generally cylindrical. It is possible to demonstrate that, with a multipole expansion of the wakefields, the dominant term in the longitudinal wakefield depends only on the distance $z$ between the two charges, while the dominant one in the transverse wakefield, although still a function of the distance $z$, is also linear with the transverse position of the source charge $r_0$. If we then divide the transverse wakefield by $r_0$, we obtain the transverse dipole wakefield, that is the transverse wake per unit of transverse displacement, depending only on $z$.

*Longitudinal wakefield* [V/C]:     $w_{/\!/}(z) = -\dfrac{U}{q_0 q}$ ;     (4)

*Transverse dipole wakefield* [V/C·m]: $\qquad w_\perp(z) = \frac{1}{r_0} \frac{M}{q_0 q}$. (5)

The minus sign in the definition of the longitudinal wakefield means that the test charge loses energy when the wake is positive. A positive transverse wake means that the transverse force is defocusing. The wakefields are properties of the vacuum chamber and the beam environment, but they are independent of the beam parameters (bunch size, bunch length, etc.).

To study the effect of wakefields on the beam dynamics, it is convenient to distinguish between short-range wakefields, which are generated by the particles at the head of the bunch affecting trailing particles in the same bunch, and those that influence the multibunch (or multiturn) beam dynamics, which are generally resonant modes trapped inside a structure, and are called long-range wakefields.

As a first example of wakefields, let us consider the longitudinal wakefield of 'space charge'. Even if in the ultrarelativistic limit with $\gamma \to \infty$, and there is no space charge effect, we can still define a wakefield by considering a moderately relativistic beam with $\gamma \gg 1$ but not infinite. It turns out that the space charge forces can fit into the definition of a wakefield, and when that is done we find that the wake depends on the beam properties such as the transverse beam radius $a$ and the beam energy $\gamma$. In Appendix A we show an example of such an interpretation. Let us consider here a relativistic beam with cylindrical symmetry and uniform transverse distribution of radius $a$. The longitudinal force acting on a charge $q$ of the beam travelling inside a cylindrical pipe of radius $b$ is given by [13]

$$F_\parallel(r,z) = \frac{-q}{4\pi\varepsilon_0 \gamma^2}\left(1 - \frac{r^2}{a^2} + 2\ln\frac{b}{a}\right)\frac{\partial \lambda(z)}{\partial z}, \qquad (6)$$

with $\lambda(z)$ the longitudinal distribution ($z > 0$ at the bunch head). Note that, since the space charge forces move together with the beam, they are constant along the accelerator if the beam pipe cross-section remains constant. We can therefore define the longitudinal wakefield per unit length [V/C·m]. To obtain the longitudinal wakefield of a piece of pipe, we just multiply by the pipe length. Assuming $r \to 0$ (particle on-axis), and a charge line density given by $\lambda(z) = q_0 \delta(z)$, we obtain

$$\frac{\mathrm{d}w_\parallel(z)}{\mathrm{d}s} = \frac{1}{4\pi\varepsilon_0 \gamma^2}\left(1 + 2\ln\frac{b}{a}\right)\frac{\partial}{\partial z}\delta(z), \qquad (7)$$

which has the peculiarity of being also dependent on the beam size $a$.

Another interesting case is the longitudinal wake potential of a resonant higher-order mode (HOM) in an RF cavity, which is an example of long-range wakefield. When a charge crosses a resonant structure, as an RF cavity, it excites the fundamental mode and HOMs. Each mode can be treated as an electric RLC circuit loaded by an impulsive current, as shown in Fig. 2.

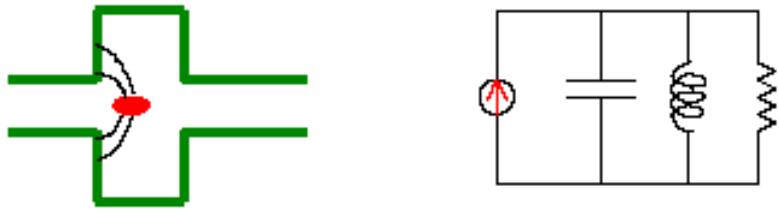

**Fig. 2:** RF cavity and the equivalent RLC parallel circuit model driven by a current generator

Just after the charge passage, the capacitor is charged with a voltage $V(0) = q_0/C$, and the longitudinal electric field is $E_z = V/l$, where $l$ is the length of the cavity. The time evolution of the electric field is then governed by the same differential equation of the voltage, which can be written as follows:

$$\ddot{V} + \frac{1}{RC}\dot{V} + \frac{1}{LC}V = \frac{1}{C}\dot{I}. \tag{8}$$

The passage of the impulsive current charges only the capacitor, which changes its potential by an amount $V_c(0)$. This potential will oscillate and decay, producing a current flow in the resistor and inductance. After the charge passage, for $t > 0$ the potential satisfies the following equation and boundary conditions:

$$\begin{aligned}&\ddot{V} + \frac{1}{RC}\dot{V} + \frac{1}{LC}V = 0, \\ &V(t=0^+) = \frac{q_0}{C} = V_0, \\ &\dot{V}(t=0^+) = \frac{\dot{q}}{C} = -\frac{I(0^+)}{C} = -\frac{V_0}{RC},\end{aligned} \tag{9}$$

which has the following solution:

$$\begin{aligned}&V(t) = V_0 e^{-\Gamma t}\left[\cos(\bar{\omega}t) - \frac{\Gamma}{\bar{\omega}}\sin(\bar{\omega}t)\right], \\ &\bar{\omega}^2 = \omega_r^2 - \Gamma^2,\end{aligned} \tag{10}$$

where $\omega_r^2 = 1/LC$ and $\Gamma = 1/2RC$. For the HOM it is also convenient to define the quality factor $Q = \omega_r/2\Gamma$, from which we can write $C = Q/R\omega_r$.

Putting $z = ct$ ($z$ is positive behind the source charge), we obtain the longitudinal wakefield shown in Fig. 3:

$$w_\parallel(z) = \frac{V(z)}{q_0} = \frac{R\omega_r}{Q} e^{-\Gamma z/c}\left[\cos(\bar{\omega}z/c) - \frac{\Gamma}{\bar{\omega}}\sin(\bar{\omega}z/c)\right]. \tag{11}$$

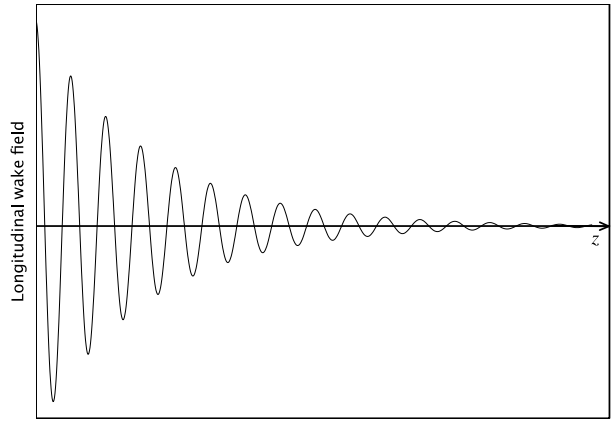

**Fig. 3:** Qualitative behaviour of a resonant mode wakefield

In an analogous way, it is possible to obtain the transverse wakefield of a HOM,

$$w_\perp(z) = \frac{R_\perp \omega_r}{Q} e^{-\Gamma z/c} \sin(\bar{\omega} z/c), \tag{12}$$

with $R_\perp$ expressed in ohm per metre.

We conclude this section by giving the longitudinal and transverse short-range wakefields of a rectangular cell, as shown in Fig. 4, under the hypothesis that the bunch length is much smaller than the pipe radius *b*. Its expression can be useful in studying the effects of the short-range wakefields of an accelerating structure in a linac.

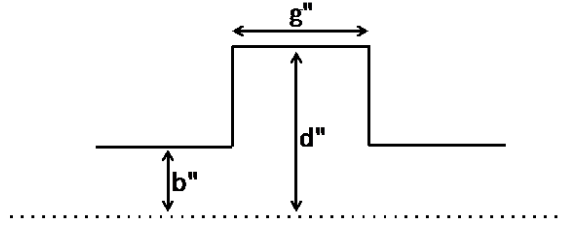

**Fig. 4:** Geometry of a single cell of a linac accelerating structure

The model considers each cell as a pillbox cavity. When a bunch reaches the edge of the cavity, the e.m. field it creates is simply the one that would occur when a plane wave passes through a hole; using this hypothesis, it is possible to use the classical diffraction theory of optics to calculate the fields [7]. If the condition $g < (d-b)^2/(2\sigma)$ is satisfied, where *g* is the cell gap, *d* is the cell radius, and $\sigma$ is the rms bunch length of a Gaussian bunch, the longitudinal and transverse wakefields can be written, respectively, as:

$$w_\parallel(z) = \frac{Z_0 c}{\sqrt{2}\pi^2 b} \sqrt{\frac{g}{z}},$$

$$w_\perp(z) = \frac{2^{3/2} Z_0 c}{\pi^2 b^3} \sqrt{gz}. \tag{13}$$

For a collection of cavities, Eqs. (13) cannot be used because the wakefields along the cells do not sum in phase and the result would be an overestimation of the effects. An asymptotic wakefield for a periodic collection of cavities of period *p*, obtained numerically at SLAC [16] and then fitted to a simple function, is used instead. Such wakefields are thus valid after a certain number of cavities given by

$$N_{cr} = \frac{b^2}{2g\left(\sigma + \dfrac{2b}{\gamma}\right)}. \tag{14}$$

Under these assumptions, the wakefields of Eqs. (13) are modified to

$$w_\parallel(z) = \frac{Z_0 c p}{\pi b^2} e^{-\sqrt{z/s_1}},$$

$$w_\perp(z) = \frac{4 Z_0 c p s_2}{\pi b^4}\left[1 - \left(1 + \sqrt{\frac{z}{s_2}}\right)e^{-\sqrt{z/s_2}}\right], \tag{15}$$

with

$$s_1 = 0.41 \frac{b^{1.8} g^{1.6}}{p^{2.4}},$$
$$s_2 = 0.17 \frac{b^{1.79} g^{0.38}}{p^{1.17}}. \tag{16}$$

## 2.2 Loss factor and beam loading theorem

A useful quantity for the effects of the longitudinal wakefield on the beam dynamics is the loss factor, defined as the normalized energy lost by the source charge $q_0$:

$$k = -\frac{U(z=0)}{q_0^2}. \tag{17}$$

For charges travelling at the velocity of light, there is the problem that the longitudinal wakefield is discontinuous at $z = 0$, as shown in Fig. 5, leading to an ambiguity in the evaluation of the loss factor. Indeed, when the source charge travels at the velocity of light, it leaves the e.m. fields in its wake, hence the reason why we call these fields 'wakefields'. Any e.m. perturbation produced by the charge cannot overtake the charge itself. This means that the longitudinal wakefield vanishes in the region $z < 0$. This property is a consequence of the *causality principle*. It is the causality that requires that the longitudinal wakefield of a charge travelling at the velocity of light to be discontinuous at the origin.

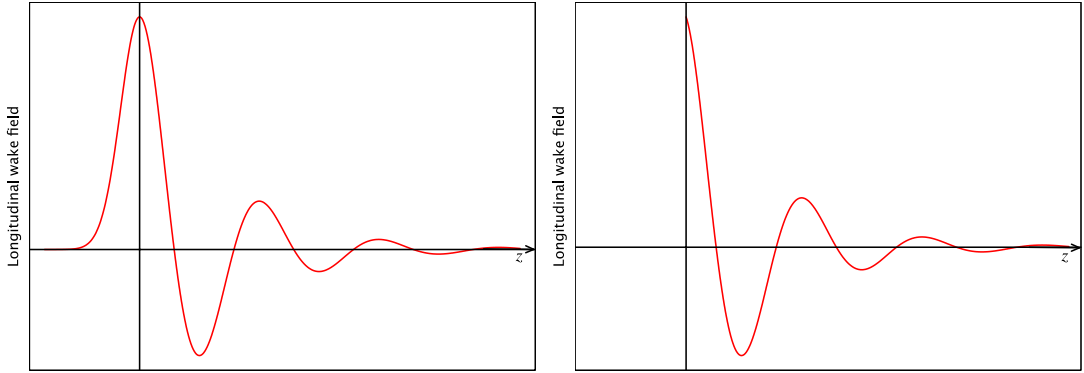

**Fig. 5:** Examples of longitudinal wakefields: left $\beta < 1$, right $\beta = 1$

The exact relationship between $k$ and $w_{//}(z \to 0)$ is, in this case, given by the beam loading theorem [17], which states that

$$k = \frac{w_{//}(z \to 0)}{2}. \tag{18}$$

As an example of verification of the beam loading theorem, let us consider the wakefield of the resonant mode given by Eq. (11). The energy lost by the charge $q_0$ loading the capacitor is $U = CV_0^2/2 = q_0^2/2C$, giving $k = 1/2C$, compared with $w_{//}(z \to 0) = 1/C$.

## 2.3 Relationship between transverse and longitudinal forces

Another important feature worth mentioning here is the differential relationship existing between longitudinal and transverse forces and the corresponding wakefields: the transverse gradient of the longitudinal force/wake is equal to the longitudinal gradient of the transverse force/wake, that is

$$\nabla_\perp F_{/\!/} = \frac{\partial}{\partial z} \boldsymbol{F}_\perp,$$
$$\nabla_\perp w_{/\!/} = \frac{\partial}{\partial z} \boldsymbol{w}_\perp. \tag{19}$$

The above relations are known as the *Panofsky–Wenzel theorem* [18].

## 2.4 Coupling impedance

The wakefields are generally used to study the beam dynamics in the time domain. If we take the equations of motion in the frequency domain, we need the Fourier transform of the wakefields. Since these quantities have units of ohm, they are called *coupling impedances*.

*Longitudinal impedance* [Ω]:

$$Z_{/\!/}(\omega) = \frac{1}{v}\int_{-\infty}^{\infty} w_{/\!/}(z) e^{i\frac{\omega z}{v}}\, dz\ ; \tag{20}$$

*Transverse dipole impedance* [Ω/m]:

$$\boldsymbol{Z}_\perp(\omega) = -\frac{i}{v}\int_{-\infty}^{\infty} \boldsymbol{w}_\perp(z) e^{i\frac{\omega z}{v}}\, dz\ . \tag{21}$$

The longitudinal coupling impedance of the space charge wake given by Eq. (7) [Ω/m] is given by

$$\frac{\partial Z_{/\!/}(\omega)}{\partial s} = \frac{1}{v}\int_{-\infty}^{\infty} \frac{\partial w_{/\!/}(z)}{\partial s} e^{i\frac{\omega z}{v}}\, dz = \frac{1+2\ln(b/a)}{v 4\pi\varepsilon_0 \gamma^2}\int_{-\infty}^{\infty} \frac{d}{dz}\delta(z) e^{i\frac{\omega z}{v}}\, dz\ ; \tag{22}$$

since $\int_{-\infty}^{\infty} \delta'(z) f(z) dz = f'(0)$, we get

$$\frac{\partial Z_{/\!/}(\omega)}{\partial s} = \frac{i\omega\, Z_0}{4\pi c \beta^2 \gamma^2}\left(1 + 2\ln\frac{b}{a}\right). \tag{23}$$

The longitudinal coupling impedance of a resonant HOM, corresponding to the Fourier transform of Eq. (11), is given by

$$Z_{/\!/}(\omega) = \frac{R}{1 + iQ\left(\dfrac{\omega_r}{\omega} - \dfrac{\omega}{\omega_r}\right)}, \tag{24}$$

where *R* is also called the shunt impedance of the longitudinal HOM. Note that the loss factor can be written as $k = \omega_r R/2Q$.

The transverse impedance obtained from Eq. (12) is given by

$$Z_\perp(\omega) = \frac{\overline{\omega}}{\omega}\frac{R_\perp}{1 + iQ\left(\dfrac{\omega_r}{\omega} - \dfrac{\omega}{\omega_r}\right)}, \tag{25}$$

where $R_\perp$ is called the transverse shunt impedance.

## 2.5 Wake potential and energy loss of a bunched distribution

When we have a bunch with total charge $q_0$ and longitudinal distribution $\lambda(z)$, such that $q_0 = \int_{-\infty}^{\infty} \lambda(z')\mathrm{d}z'$, we can obtain the amount of energy lost or gained by a single charge $q$ in the beam by using the superposition principle.

To this end, we calculate the effect on the charge by the whole bunch, as shown in Fig. 6, with the superposition principle, which gives the convolution integral:

$$U(z) = -q \int_{-\infty}^{\infty} w_{\parallel}(z'-z)\lambda(z')\mathrm{d}z'. \tag{26}$$

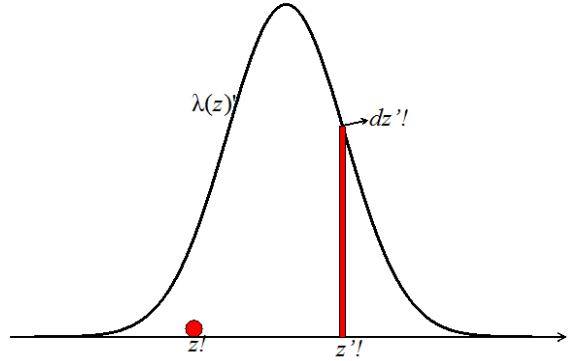

**Fig. 6:** Convolution integral for a charge distribution to obtain the energy loss of a particle due to the whole bunch.

Equation (26) allows us to define the longitudinal wake potential of a distribution:

$$W_{\parallel}(z) = -\frac{U(z)}{qq_0} = \frac{1}{q_0}\int_{-\infty}^{\infty} w_{\parallel}(z'-z)\lambda(z')\mathrm{d}z'. \tag{27}$$

The total energy lost by the bunch is computed by summing the energy loss of all the particles:

$$U_{\mathrm{bunch}} = \frac{1}{q}\int_{-\infty}^{\infty} U(z')\lambda(z')\mathrm{d}z' = -q_0 \int_{-\infty}^{\infty} W_{\parallel}(z')\lambda(z')\mathrm{d}z'. \tag{28}$$

## 3 Wakefield effects in linear accelerators

### 3.1 Energy spread

The longitudinal wake forces change the energy of individual particles depending on their position in the beam, as given by Eq. (26). As a consequence, the short-range wakefield can induce an energy spread in the beam.

For example, the energy spread induced by the space charge force in a Gaussian bunch is given by

$$\frac{\mathrm{d}U(z)}{\mathrm{d}s} = -q\int_{-\infty}^{\infty} \frac{\mathrm{d}w_{\parallel}(z'-z)}{\mathrm{d}s}\lambda(z')\mathrm{d}z' = \frac{qq_0}{4\pi\varepsilon_0\gamma^2\sqrt{2\pi}\sigma_z^3}\left(1+2\ln\frac{b}{a}\right)z\,\mathrm{e}^{-(z^2/2\sigma_z^2)}. \tag{29}$$

The bunch head gains energy ($z > 0$), and the tail loses energy. The total energy lost by the bunch, $U_{\mathrm{bunch}}$, is zero.

In a similar way, one can show that the energy loss induced by a resonant HOM inside a rectangular uniform bunch of length $l_0$ when $\Gamma \ll \bar{\omega}$ is given by

$$U(z) = \frac{-qq_0 R\omega_r}{2Q} \frac{\sin\left[\frac{\omega_r}{c}\left(\frac{l_0}{2} - z\right)\right]}{\left(\frac{\omega_r l_0}{2c}\right)}, \tag{30}$$

and the total energy loss obtained with Eq. (28) is

$$U_{bunch} = -\frac{2q_0^2 Rc^2}{\omega_r l_0^2 Q} \sin^2\left(\frac{\omega_r l_0}{2c}\right). \tag{31}$$

### 3.2 Single-bunch beam breakup: two-particle model

A beam injected off-centre in a linac, for example due to misalignments of the focusing quadrupoles, executes betatron oscillations. The bunch displacement produces a transverse wakefield in all the devices crossed during the flight, which deflects the trailing charges (single-bunch BBU), or other bunches following the first one in a multibunch regime (multibunch BBU). The first observation of BBU was made at SLAC in 1966 [19].

To understand the effect, we consider, as a first example, a simple model with only two charges, $q_1 = q_0/2$ (leading = half bunch) and $q_2 = q$ (trailing = single charge), travelling with $\beta = 1$.

The leading charge executes free betatron oscillations of the following kind:

$$y_1(s) = \hat{y}_1 \cos\left(\frac{\omega_y}{c} s\right). \tag{32}$$

The trailing charge, a distance $z$ behind, experiences over a length $L_w$ an average deflecting force that is proportional to the displacement $y_1$ and dependent on the distance $z$. From the definition of the transverse dipole wakefield, this force is given by

$$\langle F_y(z, y_1) \rangle = \frac{qq_0}{2L_w} w_\perp(z) y_1(s). \tag{33}$$

Note that $L_w$ is the length of the device for which the transverse wake has been computed. For example, in the case of a cavity cell, $L_w$ is the length of the cell. This force drives the motion of the trailing charge:

$$y_2'' + \left(\frac{\omega_y}{c}\right)^2 y_2 = \frac{qq_0 w_\perp(z)}{2E_o L_w} \hat{y}_1 \cos\left(\frac{\omega_y}{c} s\right). \tag{34}$$

This is a typical equation for a resonator driven at the resonant frequency.

The solution is given by the superposition of the 'free' and the 'forced' oscillations, which, being driven at the resonant frequency, grow linearly with $s$, as shown in Fig. 7:

$$y_2(s) = \hat{y}_2 \cos\left(\frac{\omega_y}{c} s\right) + y_2^{forced}, \tag{35}$$

$$y_2^{forced} = \frac{cqq_0 w_\perp(z) s}{4\omega_y E_o L_w} \hat{y}_1 \sin\left(\frac{\omega_y}{c} s\right). \tag{36}$$

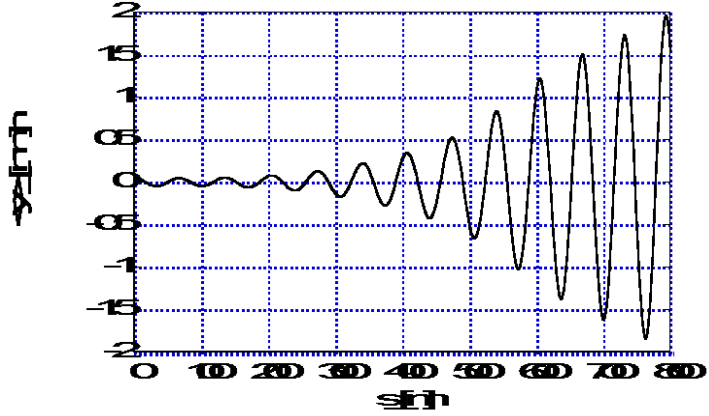

**Fig. 7:** HOMDYN [20] simulation of a typical BBU instability, 50 $\mu$m initial offset, no energy spread

At the end of a linac of length $L_L$, the oscillation amplitude will have grown by ($\hat{y}_1 = \hat{y}_2$)

$$\left(\frac{\Delta \hat{y}_2}{\hat{y}_2}\right)_{max} = \frac{cNew_\perp(z)L_L}{4\omega_y(E_o/e)L_w}. \tag{37}$$

If the transverse wake is given per cell, the relative displacement of the tail with respect to the head of the bunch depends on the number of cells. Of course, it also depends on the focusing strength through the betatron frequency $\omega_y$.

### 3.3  BNS damping

The BBU instability is quite harmful and hard to control even at high energy with strong focusing, and after careful injection and steering. A simple cure has been proposed after observing that the strong oscillation amplitude of the bunch tail is mainly due to the 'resonant' driving head. If the tail and the head move with different frequencies, this effect can be significantly reduced [15].

Let us assume that the tail oscillates with a frequency $\omega_y + \Delta\omega_y$, so that Eq. (34) becomes

$$y_2'' + \left(\frac{\omega_y + \Delta\omega_y}{c}\right)^2 y_2 = \frac{Ne^2 w_\perp(z)}{2E_o L_w}\hat{y}_1 \cos\left(\frac{\omega_y}{c}s\right), \tag{38}$$

the solution of which is given by

$$y_2(s) = \hat{y}_2 \cos\left(\frac{\omega_y + \Delta\omega_y}{c}s\right) - \frac{c^2 Ne^2 w_\perp(z)}{4\omega_y \Delta\omega_y E_o L_w}\hat{y}_1\left[\cos\left(\frac{\omega_y + \Delta\omega_y}{c}s\right) - \cos\left(\frac{\omega_y}{c}s\right)\right]. \tag{39}$$

In this case, we observe that the amplitude of the oscillation is limited and no longer grows linearly with $s$. Furthermore, by making a suitable choice for $\Delta\omega_y$, it is possible to suppress fully the oscillations of the tail. Indeed, by setting

$$\Delta\omega_y = \frac{c^2 Ne^2 w_\perp(z)}{4\omega_y E_o L_w} \tag{40}$$

if $\hat{y}_2 = \hat{y}_1$, from Eq. (39) we obtain

$$y_2(s) = \hat{y}_1 \cos\left(\frac{\omega_y}{c}s\right); \tag{41}$$

that is, the tail oscillates with the same amplitude as the head and with the same betatron frequency. This method of curing the single-bunch BBU instability is called BNS damping, after the names of the authors Balakin, Novokhatsky, and Smirnov who proposed it [15].

To have BNS damping, Eq. (40) imposes an extra focusing at the tail, which must have a higher betatron frequency than the head. This extra focusing can be obtained by: (1) using a Radiofrequency Quadrupole (RFQ), where the head and tail see a different focusing strength, (2) creating a correlated energy spread across the bunch, which, because of the chromaticity, induces a spread in the betatron frequency. An energy spread correlated with position is attainable with the external accelerating voltage or with wakefields.

In Fig. 8, we show the betatron oscillation corresponding to Fig. 7, but with a 2% energy spread.

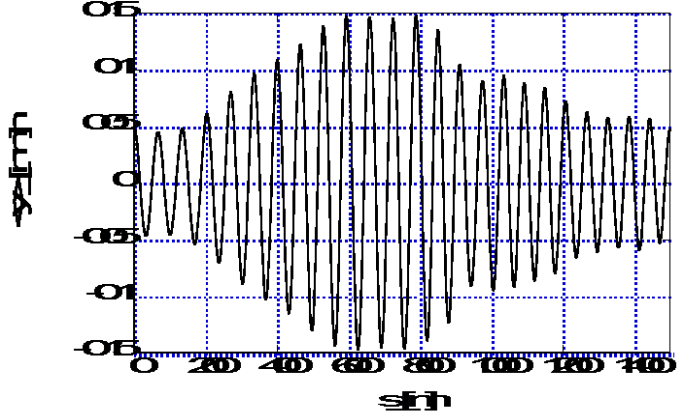

**Fig. 8:** HOMDYN simulation of a typical BNS damping, 50 $\mu$m initial offset, 2% energy spread

### 3.4 Single-bunch beam breakup: general distribution

To extend the analysis carried out in Section 3.2 to a particle distribution, we write the transverse equation of motion of a single charge $q$ with the inclusion of the transverse wakefield effects as follows [14]:

$$\frac{\partial}{\partial s}\left[\gamma(s)\frac{\partial y(z,s)}{\partial s}\right] + k_y^2(s)\gamma(s)y(z,s) = \frac{q}{m_0 c^2 L_w} \int_z^\infty y(s,z') w_\perp(z'-z) \lambda(z') \mathrm{d}z', \tag{42}$$

where $\gamma(s)$ is the relativistic parameter, which varies along the linac, and $1/k_y(s)$ is the betatron function. We recall that the integral of the longitudinal distribution function $\lambda(z)$ is the total charge of the bunch, $q_0$.

The solution of the equation in the general case is unknown. We can, however, apply a perturbation method to obtain the solution at any order in the wakefield intensity. Indeed, we write

$$y(z,s) = \sum_n y^{(n)}(z,s), \tag{43}$$

where *n* represents the *n*th-order solution. The first-order solution is found without the wakefield effect from the following equation:

$$\frac{\partial}{\partial s}\left[\gamma(s)\frac{\partial y^{(0)}(z,s)}{\partial s}\right]+k_y^2(s)\gamma(s)y^{(0)}(z,s)=0. \tag{44}$$

It is important to note that the above equation no longer depends on *z*. This means that the bunch distribution remains constant along the structure.

If the s-dependence of $\gamma(s)$ and $k_y^2(s)\gamma(s)$ is moderate, we can use the Wentzel–Kramers–Brillouin (WKB) approximation [5], and the solution of the above equation with the starting conditions $y(0)=\hat{y}$, $y'(0)=0$ is

$$y^{(0)}(s)=\sqrt{\frac{\gamma_0 k_{y0}}{\gamma(s)k_y(s)}}\hat{y}\cos[\psi(s)], \tag{45}$$

where

$$\psi(s)=\int_0^s k_y(s')\mathrm{d}s'. \tag{46}$$

Equation (45) represents the unperturbed transverse motion of the bunch in a linac.

The differential equation of the second-order solution is obtained by substituting the first-order solution (45) into the right-hand side of Eq. (42), yielding

$$\frac{\partial}{\partial s}\left[\gamma(s)\frac{\partial y^{(1)}(z,s)}{\partial s}\right]+k_y^2(s)\gamma(s)y^{(1)}(z,s)=\frac{q}{m_0 c^2 L_w}y^{(0)}(s)\int_z^\infty w_\perp(z'-z)\lambda(z')\mathrm{d}z'. \tag{47}$$

We are interested in the forced solution of the above equation that can be written in the form

$$y^{(1)}(z,s)=\hat{y}\frac{q}{m_0 c^2 L_w}\sqrt{\frac{\gamma_0 k_{y0}}{\gamma(s)k_y(s)}}G(s)\int_z^\infty w_\perp(z'-z)\lambda(z')\mathrm{d}z', \tag{48}$$

where

$$G(s)=\int_0^s \frac{1}{\gamma(s')k_y(s')}\sin[\psi(s)-\psi(s')]\cos[\psi(s')]\mathrm{d}s'$$
$$=\frac{1}{2}\int_0^s \frac{\sin[\psi(s)-2\psi(s')]}{\gamma(s')k_y(s')}\mathrm{d}s'+\frac{1}{2}\sin[\psi(s)]\int_0^s \frac{1}{\gamma(s')k_y(s')}\mathrm{d}s'. \tag{49}$$

The first integral undergoes several oscillations with *s*, and if $\gamma(s)$ and $k_y(s)$ do not vary much it is negligible, so that we can finally write

$$y^{(1)}(z,s)=\hat{y}\frac{q}{2m_0 c^2 L_w}\sqrt{\frac{\gamma_0 k_{y0}}{\gamma(s)k_y(s)}}\sin[\psi(s)]\int_0^s\frac{\mathrm{d}s'}{\gamma(s')k_y(s')}\int_z^\infty w_\perp(z'-z)\lambda(z')\mathrm{d}z'. \tag{50}$$

Note that the last integral in Eq. (50) is proportional to the transverse wake potential produced by the whole bunch, defined in a similar way to Eq. (27). This solution can then be substituted again into the right-hand side of Eq. (42) to obtain a third-order solution, and so on. If we consider constant

$\gamma(s)$ and $k_y(s)$, Eq. (50) gives the same result as for the two-particle model of Eq. (36) when we substitute $\lambda(z)$ with $q_0/2$ representing the leading half bunch affecting a trailing charge $q$.

If the BBU effect is strong, it is necessary to include higher-order terms in the perturbation expansion. Under the following assumptions:

i) rectangular bunch distribution $\lambda(z) = q_0/l_0$, $-l_0/2 < z < l_0/2$, where $l_0$ is the bunch length;

ii) monoenergetic beam;

iii) constant acceleration gradient, $dE_0/ds = $ const.;

iv) constant beta function;

v) linear wake function inside the bunch, $w_\perp(z) = w_{\perp 0} z/l_0$,

the sum of Eq. (43) can be written in terms of powers of the adimensional parameter $\eta$, also called the BBU strength,

$$\eta = \frac{qq_0}{k_y (dE_0/ds)} \frac{w_{\perp 0}}{L_w} \ln\left(\frac{\gamma_f}{\gamma_i}\right), \tag{51}$$

where $\gamma_i$ and $\gamma_f$ are the initial and final relativistic parameter, respectively.

By using the method of steepest descent [8], it is possible to obtain the asymptotic expression of $y(z,s)$ thus finding, at the end of the linac,

$$y(L_L) = y_m \sqrt{\frac{\gamma_i}{6\pi\gamma_f}} \eta^{-1/6} \exp\left[\frac{3\sqrt{3}}{4}\eta^{1/3}\right] \cos\left[k_y L_L - \frac{3}{4}\eta^{1/3} + \frac{\pi}{12}\right], \tag{52}$$

which, unlike the two-particle model and that from the first-order solution, gives a tail displacement growing exponentially with $\eta$.

### 3.5 Multibunch beam breakup

We have seen in the previous sections that when a bunch passes off-axis (due, for example, to betatron oscillations) in an axis-symmetric accelerating structure, it excites transverse wakefields which may cause the tail of the bunch to oscillate with increasing amplitude as the bunch goes along the linac. In the same way, the whole bunch may excite deflecting trapped modes in the RF cavities of the linac that may cause trailing bunches to be deflected, whether they are on-axis or not. These angular deflections are transformed into transverse displacements through the transfer matrices of the focusing system, and the displaced bunches will themselves create similar wakefields in the downstream accelerating structures of a linac. The subsequent bunches will be further deflected leading to beam blow-up. Due to the long-range wakefields, there is a coupling in the motion of the bunches that are increasingly deflected as they proceed along the linac in a process called multibunch BBU. Even if the bunches are not lost, the transverse beam emittance can be greatly increased, leading to a significant luminosity reduction.

We summarize here the analytical study of multibunch BBU performed with the formalism used in [14]. All the bunches are considered to be rigid macro-particles, such as delta-functions, separated by period $T$, and we assume all bunches to be injected with the same initial offset $x_0$. We consider the transverse equation of motion of a bunch as a whole, ignoring internal structures; the beam is therefore made of a train of bunches with the same charge ($Q_b$) evenly spaced by period $T$, which is an integer number of the RF period of the accelerating mode.

We also consider all the cells of the linac accelerating structure to be identical and with the same dipole trapped mode in each cell of length $L_w$. Rigorously, the analytical approach requires that many betatron oscillations are performed in the linac, and the BBU remains moderate within a betatron oscillation. Moreover, the theory is valid if the beam energy does not change too much in a betatron wavelength. This last hypothesis is also called adiabatic acceleration.

The transverse wakefield force experienced by the $k$th bunch, spaced $kT$ from the first bunch, depends on the transverse wakefield generated by the preceding bunches (and thus by their transverse displacement). The dipole long-range wakefield is produced by a high-order deflecting mode, identical in all the cavities of the structure, and it is described in terms of its resonant frequency $\omega_r$, the quality factor $Q$, and the dipole shunt resistance $R_\perp$ (expressed in ohm per meter).

The equations of motion are then written in terms of the Z-transform (see, e.g., Ref. [21]) since the displacement $x(kT, s)$ of the $k$th bunch at the position $s$ is a discrete function of time. The solution can be retrieved with a perturbation method, which considers its expansion as a series of the driving wakefield force.

The zeroth-order solution is given for a vanishing driving force, i.e. a pure betatron oscillation (unperturbed motion). It represents the motion of the first bunch, which is not affected by any wakefield because of the causality principle (the wakefield cannot travel ahead of the bunch itself). The $n$th-order solution is driven by the wakefield excited by the solution of the order $n-1$. Thus the first-order solution is computed from the motion of the first bunch, and it affects all the bunches, except the first one; it means that the $n$th-order solution affects only bunches of index larger then $n$. Therefore the summation of the series can be stopped at the $M$th order of a train of $M$ bunches. The $n$th-order solution in the Z-domain can be written as follows [14]:

$$x_n(z,s) = \sqrt{\frac{\gamma_0 k_{y0}}{\gamma(s) k_y(s)}} x_0 e^{i\psi(s)} \frac{a^n(s)}{i^n n!} G_n(z), \tag{53}$$

where $a(s)$ is the so-called dimensionless BBU strength given, in case of constant $k_y(s)$, by

$$a(s) = \frac{Q_b}{2 k_{y0} G} \omega_r \frac{R_\perp}{L_w Q} \ln\left[\frac{\gamma(s)}{\gamma_0}\right], \tag{54}$$

where $G$ is the accelerating gradient [V/m], and

$$G_n(z) = \frac{z}{z-1} \tilde{w}_\perp^n(z) \tag{55}$$

with

$$\tilde{w}_\perp(z) = \frac{1}{2i}\left(\frac{z}{z-z_1} - \frac{z}{z-z_2}\right) \tag{56}$$

and

$$z_{1,2} = e^{-\frac{T\omega_r}{2Q}} e^{\pm i\omega_r T}. \tag{57}$$

The inverse Z-transform of $x_n(z,s)$, i.e. $x_n(kT,s)$, can then be summed to get the transverse displacement of the $k$th bunch as

$$x(kT,s) = \sum_{n=0}^{\infty} x_n(kT,s). \tag{58}$$

We recall that the sum can be stopped at the *M*th term for a beam containing *M* bunches.

For $a(s) \ll 1$ the series expansion can be stopped at the first-order term, although, if the BBU strength parameter *a* is moderate, it is sufficient to keep only a few terms of the summation.

In the Z-domain, the *n*th-order solution, given by Eq. (53), has been determined analytically, and the same is possible with its infinite sum, but its inverse Z-transform, Eq. (58), is, in general, not possible to write in a closed analytical form. It is, however, possible to compute the exact solution for the *n*th bunch as a sum of *n* terms if the BBU instability is moderate in a betatron period. Moreover, it is possible to use an asymptotic technique, valid when the blow-up is strong, to have an expression of the transverse displacement that puts in evidence the main parameters playing an important role in the instability.

The asymptotic transverse displacement of the kth bunch, expressed in terms of the oscillation amplitude only, is [14]

$$x(kT,s) = x_\infty(s) + \sqrt{\frac{\gamma_0 k_{y0}}{\gamma(s) k_y(s)}} \frac{x_0 e^{-k\frac{T\omega_r}{2Q}}}{\sqrt{2\pi}\left(2e^{\frac{T\omega_r}{2Q}}\right)^{1/2}\left[\cosh\left(\frac{T\omega_r}{2Q}\right)-\cos(\omega_r T)\right]^{1/2}} \frac{[2a(s)k]^{1/4}}{2k} e^{[2a(s)k]^{1/2}},$$

(59)

where $x_\infty(s)$ is the steady state solution that is reached when long (rigorously infinite) trains of bunches are accelerated.

In Fig. 9, we show a comparison between the analytical solution obtained by numerically solving Eq. (58) and a simple tracking code that considers the bunches in the train as rigid macro-particles, but which can also take into account the contribution of several resonant modes, along with different initial offsets and displacements of the bunches. The parameters used for the calculations are given in Table 1. They refer to a C-band linac with the BBU effect produced by a HOM. In the vertical axis the normalized transverse position, evaluated at the exit of the linac, is defined as follows:

$$\frac{x(kT,s)}{x_0}\sqrt{\frac{\gamma(s) k_y(s)}{\gamma_0 k_{y0}}}.$$

(60)

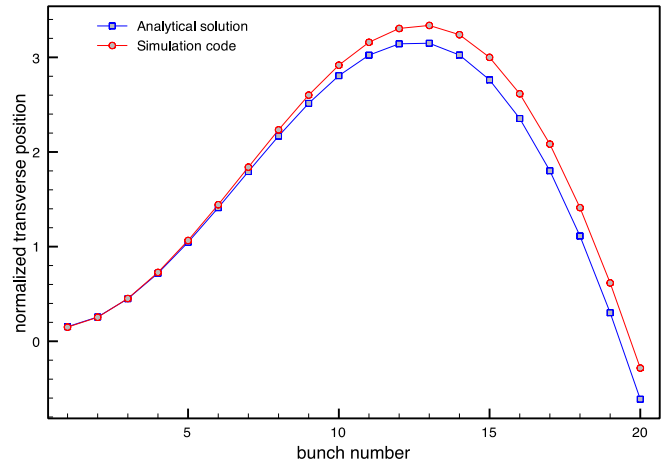

**Fig. 9:** Normalized transverse position as a function of the bunch number: comparison between the analytical solution and a tracking code.

**Table 1:** Beam parameters used for comparing the analytical solution of multibunch BBU with the results of a tracking code.

| | |
|---|---|
| Linac length | 30 m |
| Initial energy | 80 MeV |
| Energy gradient | 30 MeV/m |
| Betatron function, $1/k_y$ | 1 m |
| Bunch spacing, $T$ | 15 ns |
| Bunch charge | 1 nC |
| HOM resonant frequency, $f_r$ | 8.4 GHz |
| HOM transverse impedance, $R_\perp$ | 50 MΩ/m |
| HOM quality factor | 11 000 |
| Cell length | 17.5 cm |

From Eq. (53), we see that one possible way to reduce the BBU instability is to act on the dimensionless BBU strength given by Eq. (54). For example, we can reduce the bunch charge $Q_b$ or the betatron function, i.e. increase the focusing strength. A better approach is to remove the source of the instability by damping the transverse dipole mode, for example with an improved electromagnetic design of the accelerating cells.

The other main approach to BBU instability suppression is to detune the cell frequencies to introduce a spread in the resonance frequency of the dangerous mode so that it will no longer be excited coherently by the beam. Indeed, by properly detuning each cell, a damping of the BBU instability is produced via a decoherence of the various cell wakefields. It has been demonstrated [22] that a Gaussian distribution of the cell frequencies, which provides a rapid drop in the wakefield for a given total frequency spread, would be optimal. The analytical approach to determine the effectiveness of this detuning technique for the BBU multibunch instability can be found in Ref. [14], where it is also shown that the damping increases with the amplitude of the frequency spread.

## Appendix A: Power radiated by a bunch passing through a taper

In the case of a uniform charge distribution, and $\gamma \to \infty$, the electric field lines of a beam passing inside a perfectly conducting circular pipe are perpendicular to the direction of motion and travel together with the charge [9], as shown in Fig. A.1. In other words, the field map does not change during the charge flight, as long as the trajectory is parallel to the pipe axis. Under this condition, the transverse field's intensity can be computed as in the static case, applying Gauss's and Ampere's laws:

$$\int_S \varepsilon_0 \mathbf{E} \cdot \mathbf{n} \, dS = \int_V \rho \, dV, \qquad \oint \mathbf{B} \cdot d\mathbf{l} = \mu_0 \int_S \mathbf{J} \cdot \mathbf{n} \, dS \qquad (A.1)$$

Let us consider a cylindrical beam of radius $a$ and current $I$, with uniform charge density $\rho = I/\pi a^2 v$ and current density $J = I/\pi a^2$, propagating with relativistic speed $v = \beta c$ along the axis of a cylindrical perfectly conducting pipe of radius $b$, as shown in Fig. A.1.

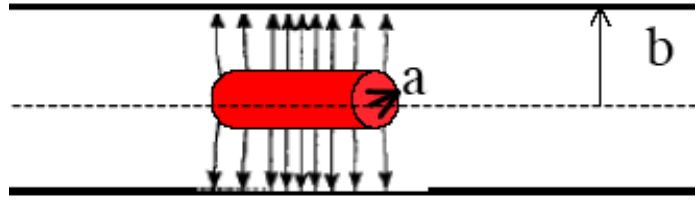

**Fig. A.1:** Cylindrical bunch of radius $a$ propagating inside a cylindrical perfectly conducting pipe of radius $b$

By applying Eqs. (A.1), one obtains for the radial component of the electric field:

$$E_r = \frac{I}{2\pi\varepsilon_0 a^2 v} r \quad \text{for} \quad r \leq a,$$

$$E_r = \frac{I}{2\pi\varepsilon_0 v} \frac{1}{r} \quad \text{for} \quad r > a,$$

and the relation $B_\vartheta = \dfrac{\beta}{c} E_r$ holds.

The electrostatic potential satisfying the boundary condition $\varphi(b) = 0$ is given by

$$\varphi(r,z) = \int_r^b E_r(r',z)\,dr' = \begin{cases} \dfrac{I}{4\pi\varepsilon_0 v}\left(1 + 2\ln\dfrac{b}{a} - \dfrac{r^2}{a^2}\right) & \text{for } r \leq a \\[6pt] \dfrac{I}{2\pi\varepsilon_0 v} \ln\dfrac{b}{r} & \text{for } a \leq r \leq b. \end{cases}$$

How can a perturbation of the boundary conditions affect the beam dynamics? Let us consider the following example: a smooth transition of length $L$ (taper) from a beam pipe of radius $b$ to a larger beam pipe of radius $d$ is experienced by the beam [9]. To satisfy the boundary condition of a perfectly conducting pipe also in the tapered region, the field lines are bent, as shown in Fig. A.2. Therefore there must be a longitudinal electric field $E_z(r,z)$ in the transition region.

A test particle moving outside the beam charge distribution will experience along the transition of length $L$ a voltage difference given by (e.g. see Ref. [21]):

$$V = -\int_z^{z+L} E_z(r,z')\,dz' = -[\varphi(r,z+L) - \varphi(r,z)] = -\frac{I}{2\pi\varepsilon_0 v}\ln\frac{d}{b},$$

which is decelerating if $d > b$. The power lost by the beam to sustain the induced voltage is given by

$$P_{lost} = VI = \frac{I^2}{2\pi\varepsilon_0 v}\ln\frac{d}{b}. \tag{A.2}$$

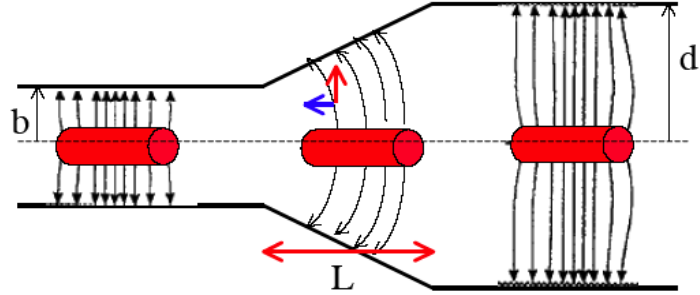

**Fig. A.2:** Smooth transition of length $L$ (taper) from a beam pipe of radius $b$ to a larger beam pipe of radius $d$

This means that for $d > b$ the power is deposited into the energy of the fields: moving from left to right of the transition, the beam induces the fields in the additional space around the bunch (i.e. in the region $b < r < d$, $0 < z < l_0$) at the expense of the only available energy source, which is the kinetic energy of the beam itself.

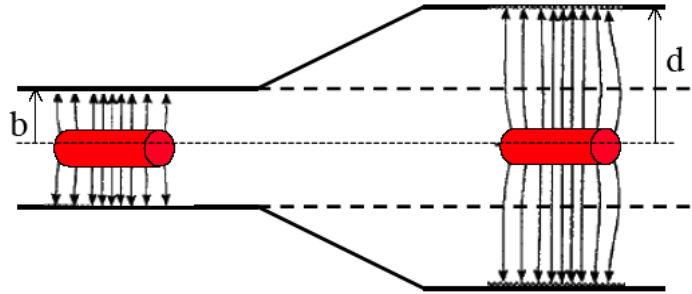

**Fig. A.3:** During the beam propagation in the taper, additional e.m. power flow is required to fill the new available space.

To verify this interpretation, let us now compute the e.m. power radiated by the beam to fill the additional space available around the bunch, as shown in Fig. A.3. On integrating the Poynting vector through the surface $\Delta S = \pi(d^2 - b^2)$ representing the additional power passing through the right part of the beam pipe, one obtains

$$P_{em} = \int_{\Delta S}\left(\frac{1}{\mu}\mathbf{E}\times\mathbf{B}\right)\cdot\mathbf{n}\,dS = \int_b^d \frac{E_r B_\vartheta}{\mu}2\pi r\,dr = \frac{I^2}{2\pi\varepsilon_0 v}\ln\frac{d}{b},$$

which is exactly the same expression as in Eq. (A.2). Note that if $d < b$ the beam gains energy. If $d \to \infty$, the power goes to infinity. Such an unphysical result is nevertheless consistent with the original assumption of an infinite energy beam ($\gamma \to \infty$).